\documentclass[conference]{IEEEtran}
\IEEEoverridecommandlockouts

\usepackage{balance}
\usepackage{graphicx}
\usepackage{textcomp}
\usepackage{todonotes}
\usepackage{import}
\usepackage{colortbl}
\usepackage{tcolorbox}

\usepackage{graphicx}
\usepackage{soul}
\usepackage{pifont}
\usepackage{enumitem}
\usepackage{url}
\usepackage{footmisc}
\usepackage{blindtext}
\usepackage{dirtytalk} 
\usepackage{cleveref}
\usepackage{multirow}
\usepackage{graphicx}
\usepackage{arydshln}
\graphicspath{ {./Figure/} }

\makeatletter
\newcommand{\linebreakand}{%
  \end{@IEEEauthorhalign}
  \hfill\mbox{}\par
  \mbox{}\hfill\begin{@IEEEauthorhalign}
}
\makeatother

\newcommand{\toolName}{\textsc{Catto}\xspace}
\newcommand{\catto}{\textsc{Catto core}\xspace}
\newcommand{\cattoIJ}{\textsc{Catto IntelliJ IDEA}\xspace}
\newcommand{\soot}{\textsc{Soot}\xspace}
\newcommand{\smugtool}{\textsc{SMUG}\xspace}

\newcommand{\ie}{i.e.,\xspace}
\newcommand{\eg}{e.g.,\xspace}

\newcommand{\etal}{et al.\xspace}

\newcommand*\circled[1]{\tikz[baseline=(char.base)]{
            \node[shape=circle,draw,inner sep=2pt] (char) {#1};}}

\begin{document}

\title{CATTO: Just-in-time Test Case Selection and Execution}

\author{\IEEEauthorblockN{
Dario Amoroso d'Aragona\IEEEauthorrefmark{1}, 
Fabiano Pecorelli\IEEEauthorrefmark{1},
Simone Romano\IEEEauthorrefmark{2},
Giuseppe Scanniello\IEEEauthorrefmark{2},
}
\IEEEauthorblockN{
Maria Teresa Baldassarre\IEEEauthorrefmark{3}, 
Andrea Janes\IEEEauthorrefmark{4}, 
Valentina Lenarduzzi\IEEEauthorrefmark{5}}

\IEEEauthorblockA{
dario.amorosodaragona@tuni.fi, 
fabiano.pecorelli@tuni.fi,
siromano@unisa.it,
gscanniello@unisa.it
}
\IEEEauthorblockA{
teresa.baldassarre@uniba.it, 
andrea.janes@unibz.it, 
valentina.lenarduzzi@oulu.fi}

\IEEEauthorblockA{
    \IEEEauthorrefmark{1}Tampere University, Finland, 
    \IEEEauthorrefmark{2}University of Salerno, Italy,
    \IEEEauthorrefmark{3}University of Bari, Italy}
\IEEEauthorblockA{
    \IEEEauthorrefmark{4}Free University of Bozen-Bolzano, Italy, 
    \IEEEauthorrefmark{5}University of Oulu, Finland}
}














\maketitle
\thispagestyle{plain}
\pagestyle{plain}
\begin{abstract}
Regression testing wants to prevent that errors, which have already been corrected once, creep back into a system that has been updated. A na\"ive approach consists of re-running the entire test suite (TS) against the changed version of the software under test (SUT). However, this might result in a time- and resource-consuming process; \eg when dealing with large and/or complex SUTs and TSs. To avoid this problem, Test Case Selection (TCS) approaches can be used. This kind of approaches build a temporary TS comprising only those test cases (TCs) that are relevant to the changes made to the SUT, so avoiding executing unnecessary TCs. In this paper, we introduce \toolName (Commit Adaptive Tool for Test suite Optimization), a tool implementing a TCS strategy for SUTs written in Java as well as a wrapper to allow developers to use \toolName within IntelliJ IDEA and to execute \toolName just-in-time before committing changes to the repository. We conducted a preliminary evaluation of \toolName on seven open-source Java projects to evaluate the reduction of the test-suite size, the loss of fault-revealing TCs, and the loss of fault-detection capability. The results suggest that \toolName can be of help to developers when performing TCS. The video demo and the documentation of the tool is available at: \url{https://catto-tool.github.io/}

\end{abstract}

\begin{IEEEkeywords}
Software testing, test case selection, regression testing 
\end{IEEEkeywords}

\section{Introduction}
Regression testing wants to prevent that errors, which have already been corrected once, creep back into a system that has been updated~\cite{Yoo:2010}. A na\"ive approach, namely \textit{Retest-all}, consists of re-running the entire test suite (TS) against the changed version of the software under test (SUT)~\cite{rothermel1997safe}. The problem with Retest-all is that the re-execution of the entire TS might result in a time- and resource-consuming process, especially when the SUT and its TS grow in size and/or complexity. Moreover, there are Agile development practices, like test-driven development~\cite{Erdogmus:2010}, which leverage continuous regression testing, requiring the developer to execute the TS several times during a development session. If the execution of the TS demands too much time or too many resources, the developer is likely not to execute regression tests as many times as the Agile development practice would require. To tackle this problem, researchers have devised several strategies, which can be grouped into three main groups: test suite minimization (or reduction), Test Case Selection (TCS), and test case prioritization~\cite{Yoo:2010}. Both test suite minimization and TCS strategies seek to reduce the size of the TS that will be re-executed against the SUT. To that end, test suite minimization strategies remove redundant/obsolete test cases (TCs), either temporarily or permanently, from the TS. On the other hand, TCS strategies temporarily remove TCs that are not modification-aware. In other words, TCS strategies build a temporarily TS by selecting a subset of TCs (from the original TS) that are relevant to the changes made to the SUT, so avoiding executing those TC that do not exercise the changed parts. Finally, TC prioritisation strategies concern the identification of an ``ideal'' ordering of TCs maximizing some desirable properties (\eg early fault detection).

In this paper, we introduce \toolName (Commit Adaptive Tool for Test suite Optimization), a tool implementing a TCS strategy for SUTs written in Java that selects TCs to be re-executed by comparing the call graphs of the two versions of the SUT (\ie the versions before and after some changes are made to the SUT). Unlike other TCS tools (\eg SPIRITuS~\cite{romano2018spiritus}, Pythia~\cite{vokolos1997pythia}, or TestTube~\cite{chen1994testtube}), \toolName does not require code coverage information but determines a link between source code and corresponding test cases using a generated call graph (see Sect. \ref{sec:architecture}). It can be therefore used in application contexts where code coverage information (\eg statements covered by TCs) is not available. Moreover, to allow developers to execute \toolName just-in-time before committing changes to the repository, we implemented a plugin for IntelliJ IDEA \cite{IntelliJ} that wraps its functionalities. 

\toolName was specifically developed with the intention to focus on fast feedback during the development cycle, preferring technologies that are not resource intensive, also accepting a lower accuracy of the tool. We conducted a preliminary evaluation of \toolName on seven open-source Java projects to evaluate the reduction of the test-suite size, the loss of fault-revealing TCs, and the loss of fault-detection capability.
 



\section{CATTO Components}
\toolName consists of two main components: \catto and \cattoIJ. \catto contains the application logic, \cattoIJ is a plugin for IntelliJ IDEA to provide the features of \catto during development. 

\begin{figure}[h!]
\begin{center} 
\includegraphics[width=\linewidth]{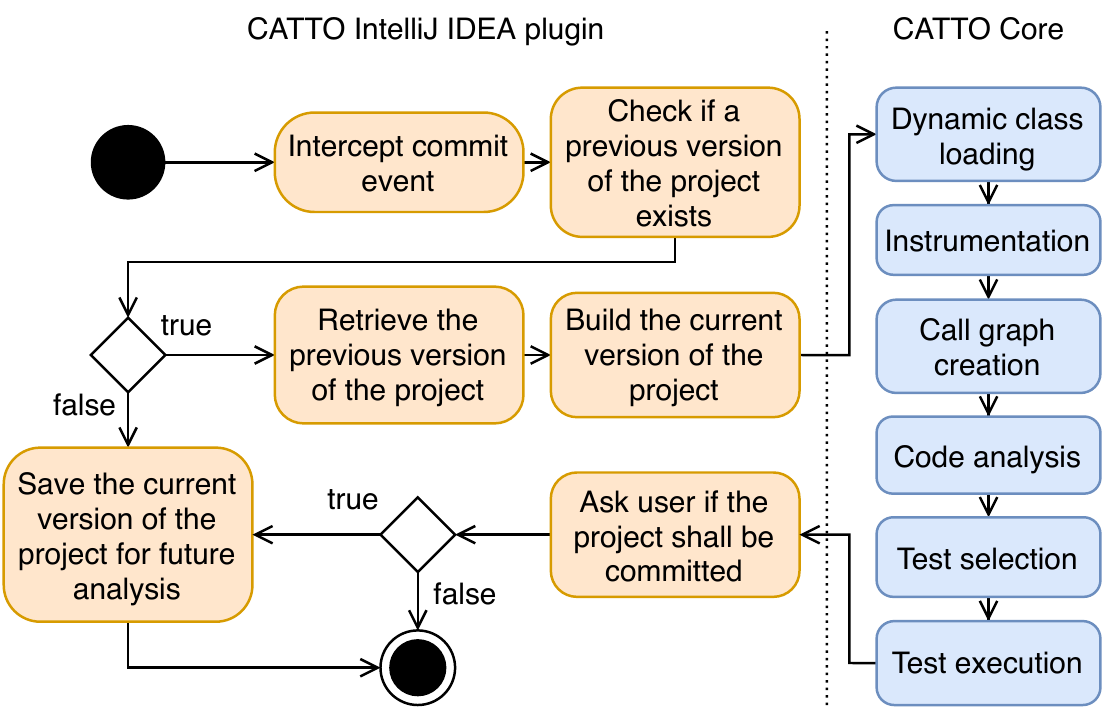}
\caption{Activity diagram of \toolName and the interaction between its parts}
\label{fig:activity-diagram} 
\end{center}
\end{figure}

As shown in Fig.~\ref{fig:activity-diagram} (on the left hand side) \catto performs six sequential steps, described as follows.

\begin{enumerate}
    \item \textbf{Dynamic classes loading}: the tool loads the classes of the current and previous version of the project and their dependencies. We rely on \soot, a third-party component, to transform byte-code in an intermediate representation (called ``JIMPLE'' \cite{Vallee-rai:2004}), necessary to have a normalized representation of the code that is easier to compare. 
    \item \textbf{Instrumentation}: Once all classes are loaded, \catto dynamically creates a Java test class for each actual test class, adding to each test method a call to the fixture methods if they are declared in the class itself or in a parent class (\eg \textit{setUp()} and \textit{tearDown()}). This step is required to ensure the creation of a reliable call graph.
    \item \textbf{Call graph creation}: Once the fake test classes are created, \catto uses these as a starting point for the generation of the call graph of the current version of the project. The generation of the call graph starts from the test methods and goes through the production code, mapping in this way each production method to its corresponding test methods. 
    \item \textbf{Code Analysis}: Then, \catto analyzes the two versions of the project searching for code changes and marking all the methods and class modified. \catto marks a method or class as changed according to the operations described in Tab. \ref{tab:criteria}.
    \item \textbf{Test Case Selection}: The methods marked in the previous step are used by \catto to select the test methods to execute. For each marked method, the corresponding test methods are selected. 
    Tab. \ref{tab:criteria} describes all the criteria the tool adopts to perform TCS.
    \item \textbf{Test Case Execution}: Finally, when the test methods are selected, \catto executes them and displays the results, as well as the error stack trace in case of failure.  
\end{enumerate}
%
%
\begin{table}[]
\centering
\footnotesize
\setlength{\tabcolsep}{4pt}
\caption{Selection criteria description.}
\label{tab:criteria}
\begin{tabular}{p{3.8cm}|p{4.7cm}} \hline 
\textbf{Operation}                          & \textbf{Selected Test Case} \\ \hline
\textit{Production Method}                           & \\
-- added or modified                        & The respective test methods \\
-- deleted                                  & Test methods that covered it \\ 
-- deleted in a hierarchy                   & All tests methods covering the actual production methods in hierarchy \\ \hdashline[1pt/1pt]
\textit{Constructor of a production class}          & \\
-- added, modified or deleted               & All test methods covering the production methods in that class \\ \hdashline[1pt/1pt]
\textit{Static field of a production class}          & \\
-- added, modified, or deleted              & All test methods covering the production methods in that class \\ \hdashline[1pt/1pt]
\textit{Test Method}                                 & \\
-- added                                    & The new test method                                                   \\
-- modified                                 & The modified test method                                              \\ \hdashline[1pt/1pt]
\textit{Static field, constructor or fixture method of a test class} & \\
-- modified                                 & All test methods in that test class                                   \\ \hline
\end{tabular}
\end{table}
The component \cattoIJ extends the functionality of \catto integrating it in IntelliJ IDEA. This integration not only provides a user interface of \catto within a development environment, but also (1)~catches the commit event and prepares \catto to execute the analysis before the code is committed; (2)~retrieves the previous version of the project (the version of the project at the time of the last executed commit); (3)~builds the current version of the project; (4)~runs \catto; (5)~returns the test execution summary to the user and asks them whether to proceed with the commit or not; (6)~saves the committed version of the project in a hidden folder for future analyses. Figure~\ref{fig:activity-diagram} (left hand side) shows the order in which these steps take place and how the \catto steps are integrated within the \cattoIJ to perform TCS. 

x\begin{figure}[b] 
\begin{center} 
\includegraphics[width=\linewidth]{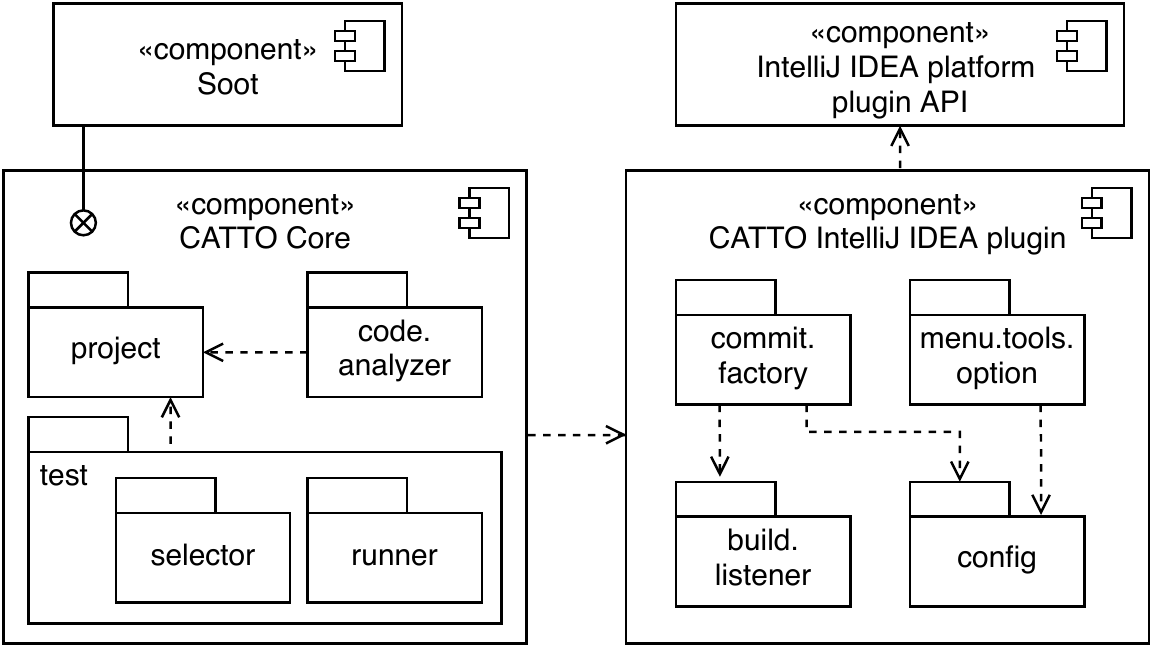}
\caption{\toolName architecture at the package level}
\label{fig:architecture-diagram} 
\end{center}
\end{figure}

\section{The CATTO Architecture}
\label{sec:architecture}

\Cref{fig:architecture-diagram} depicts the architecture of our tool in form of a UML component diagram also depicting the relevant packages. \catto relies on \soot as a third-party component and \cattoIJ exploits the \textit{IntelliJ IDEA Platform SDK API}. Due to space limitations, we discuss only the architecture of \catto, being it the core of the system. 

\catto is composed by four main packages: $project$, $code.analyzer$, $test.selector$, $test.runner$.

The $project$ package manages a single version of the project and contains all the information about it  (\eg dependencies). It allows distinguishing between the current and the previous version of a project, setting up \soot differently depending on the case, in particular creating the call graph only for the current version of the project. 

The $code.analyzer$ package provides change analysis features between two different code versions. To allow a truthful comparison between two versions of the same method, \toolName checks the name of the package and the header of the method. When the package name and the header are equal, \toolName compares the body of the methods marking the method as modified if there are some changes in the body. When the header of the method and the package are different, the method is marked as new if does not belong to any class of the previous version, otherwise, it is marked as deleted. By using the byte-code and transforming this into a \soot normalized intermediate representation, \toolName is also able to ignore cosmetic changes in the code.


The $test.selector$ package analyzes the call graph to find the tests to select. To minimize the required time to analyze the call graph, \toolName utilizes multi-threading and performs the search starting from the marked methods, going up the graph until a test method is found. For each test method, the corresponding sub-call graph is analyzed; the above mentioned search strategy allows the tool to analyze only those sub-call graphs containing a marked method. 

The $test.runner$ package allows running the selected test methods. To allow Junit to launch the test methods, the project and its dependencies are loaded dynamically and added to the JAVA classpath. The use of JUnit \cite{JUnit} to execute the test methods ensures that all and only the necessary fixture methods (\eg methods tagged with \texttt{@Before} or \texttt{@After}) are executed before and after the selected test methods.

\section{Typical Usage Scenario}

\begin{figure*}[h] 
\begin{center} 
\includegraphics[width=0.76\linewidth]{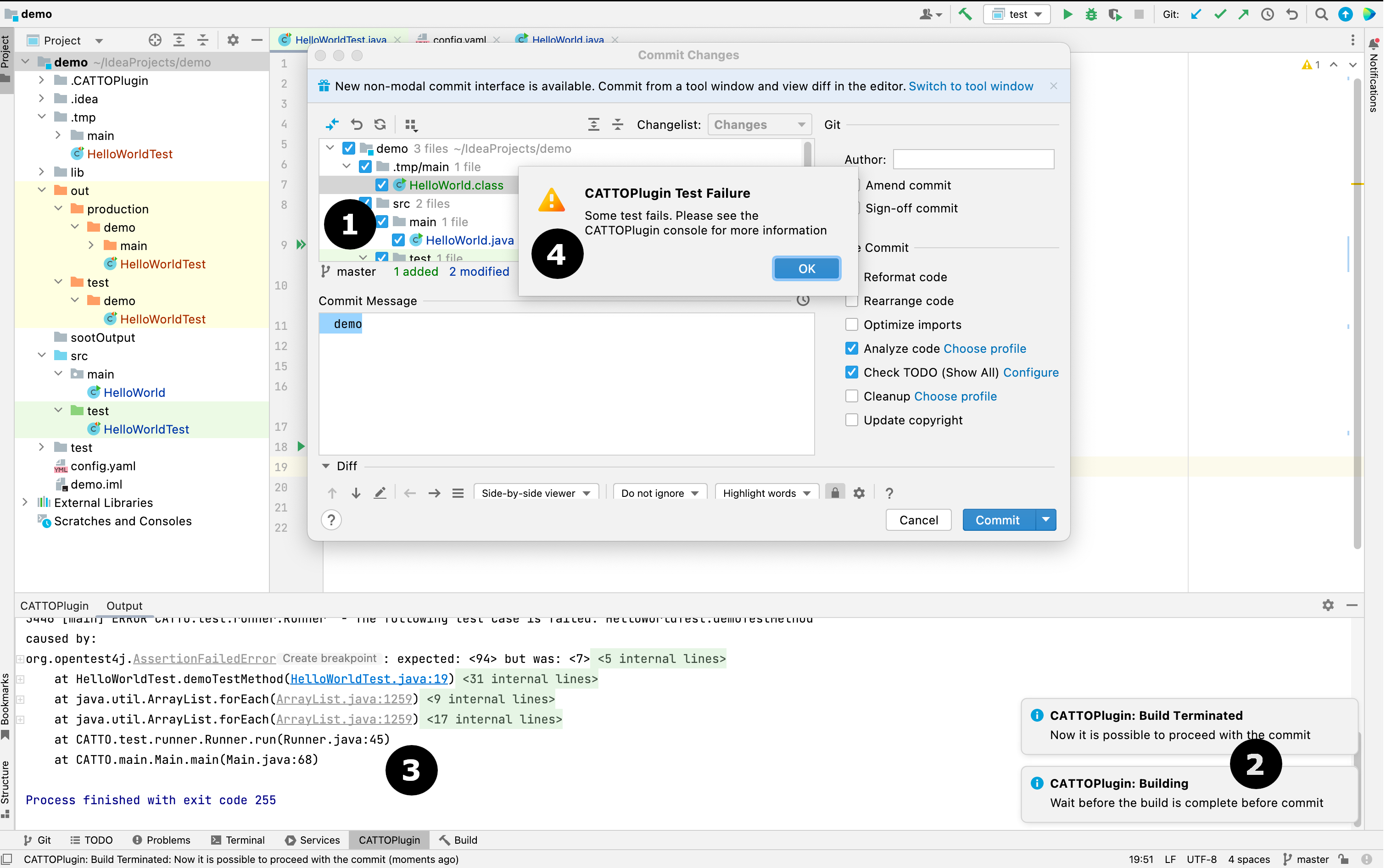}
\caption{Screenshot of an interaction with the \cattoIJ}
\label{fig:screenshot} 
\end{center}
\end{figure*}

This section describes how developers can use \toolName in a typical scenario.
Lets suppose Alice, a developer, is adding a new feature to a system. After concluding the work, she will commit the changes. At this point, \toolName intercepts the commit event, selects and runs only the test methods that cover the newly implemented methods (see step {\footnotesize\circled{1}} and {\footnotesize\circled{2}} in \Cref{fig:screenshot}) and shows the results of the executed test methods in the console (see step {\footnotesize\circled{3}} in \Cref{fig:screenshot}). If some tests fail, \toolName notifies the failure to Alice (see step {\footnotesize\circled{4}} in \Cref{fig:screenshot}), thus, she can choose whether to commit anyway or to fix the issue(s) before proceeding with the commit.

Now suppose that Alice continuous her work and after a while somebody notices a bug in the new feature. Alice finds out what has to be changed, updates the code, adds testing code to cover it and commits the changes. \toolName, intercepts the commit, selects the new test methods and the test methods that cover the changes in the code, runs these, and, as before, shows the output in the console and notifies Alice about the results of the regression testing. 

In summary, \toolName encourages Alice to practice continuous testing intercepting git \cite{GIT} commits and visualizing tests results directly within the IDE.

\section{Tool Validation}
The validation was performed using seven Java open source systems (denoted below as S1--S7): 
\textit{Apache Commons IO} \cite{S1}, 
\textit{Apache Commons Beanutils} \cite{S2}, 
\textit{Apache Commons Codec} \cite{S3}, 
\textit{DBCP} \cite{S4}, 
\textit{JXPath} \cite{S5}, 
\textit{JFreeChart} \cite{S6} and
\textit{JGap} \cite{S7}. 

We selected these systems because they are Java projects, are open-source, belong to different domains, and have non-trivial test suites. 
The validation was performed using \smugtool \cite{romano2017smug}, a mutation generator tool. To simulate source code changes, for each of the seven considered projects, we created 30 mutated versions using \smugtool (using the default configuration). 
While creating the mutated versions, \smugtool executes the test methods tracking the failures. We validated \toolName by considering the original version as the previous version and the $n{th}-$mutated version as the current version. As an outcome, we expected that all the failed test methods would be selected. 
For each pair (original and mutated version) we calculated the following metrics, referring to the fault revealing selected test methods as $X$ and the fault revealing test methods as $Y$:

\begin{itemize}
    \item \textbf{Test Suite Reduction (TSR)}: the percentage of reduction of the test suite, calculated as the proportion between the number of test methods in the original test suite and the number of test method in the selected test suite. The higher the value, the smaller the selected test suite. Therefore, the most desirable value is 1. 
    \item \textbf{Inclusiveness. (I)}: the capability to select all the fault-revealing TCs. 
    $I = \frac{X}{Y}, \;if\; Y \neq 0,1 \;if\; Y = 0$. The most desirable value is 1. 
    \item \textbf{Reduction in Fault Detection Capability (RFDC)}: the loss of capacity of the selected test methods in revealing faults. 
    $RFDC = 1 - \frac{X}{Y} \;if\; Y \neq 0,1\;if\; Y = 0$. The most desirable value is 0.
\end{itemize}
For each system we calculated the average value of each metric obtained comparing the original and each mutated version.

The results reported in \Cref{tab:validation} show that in two systems the mean of the TSR is more than $0.85$, in one is more than $0.6$, in two is more than $0.3$ and in two is below $0.2$. For the Inclusiveness, in all the systems the mean is more than $0.7$, with two cases where is more than $0.95$. Finally, the mean of RFDC, for all the systems, is between $0$ and $0.15$. 
\begin{table}[h]
\centering
\footnotesize
\caption{Validation results}
\label{tab:validation}
\begin{tabular}{c|cclllll} \hline
              & \textbf{S1} & \textbf{S2}   & \textbf{S3}   & \textbf{S4}   & \textbf{S5}   & \textbf{S6}   & \textbf{S7} \\ \hline 
\textbf{TSR}  & 0,63        & 0,31          & 0,90          & 0,86          & 0,25          & 0,40          & 0,09 \\
\textbf{I}    & 0,88        & 0,84          & 0,87          & 0,78          & 0,71          & 0,97          & 0,96 \\
\textbf{RFDC} & 0,03        & 0,01          & 0             & 0             & 0,11          & 0             & 0,14 \\\hline
\end{tabular}
\end{table}%

\section{Related Work}

Researchers have proposed several Test Case Selection (TCS) approaches based on various techniques. Integer programming~\cite{fischer1977test,fischer1981methodology}, data-flow analysis~\cite{harrold1988incremental,taha1989approach,gupta1992approach,wong1997study}, graph walking~\cite{rothermel1993safe,rothermel1994selecting,rothermel1997safe,rothermel2000regression}, textual difference and information retrieval~\cite{romano2018spiritus,vokolos1998empirical,vokolos1997pythia}, modification detection~\cite{chen1994testtube}, and firewall~\cite{Ekstazi:2015,Gligoric:2015} are just some of the techniques that the TCS approaches available in the literature rely on.

Fischer \etal~\cite{fischer1977test,fischer1981methodology} introduced one of the earliest TCS approaches. This approach used integer programming to represent the TCS problem in the context of Fortran code.
Harrold and Soffa~\cite{harrold1988incremental} applied data-flow analysis as a test case selection criterion. Similarly, Taha \etal~\cite{taha1989approach} provided a test case selection framework based on an incremental data-flow analysis method. Gupta \etal~\cite{gupta1992approach} used program slicing techniques to find definition-use pairs that were impacted by a code change. Wong \etal~\cite{wong1997study} integrated a data-flow selection strategy with coverage-based minimisation and prioritisation.

Rothermel and Harrold proposed TCS approaches based on graph walking. First, they focused on graph walking of control dependence graphs~\cite{rothermel1993safe}. Later, they extended their approach by using program dependence graphs for intra-procedural selection, and system dependence graphs for inter-procedural selection~\cite{rothermel1994selecting}. They further extended their work by relying on control flow~\cite{rothermel1997safe} and inter-procedural control flow~graphs~\cite{rothermel2000regression}.


Other approaches rely on the use of textual difference and information retrieval. Volkolos and Frankl~\cite{vokolos1997pythia,vokolos1998empirical} presented Pythia, a TCS approach based on the textual difference between the source code of the two versions of the SUT. Romano \etal~\cite{romano2018spiritus} presented SPIRITuS, an information retrieval-based TCS approach that uses method code coverage information and a vector space model to select test cases to be run.
Chen \etal~\cite{chen1994testtube} introduced TestTube, a testing framework that selects test cases based on modification detection. TestTube tracks the execution of test cases to create links between test cases and the program components it exercises.
Finally, Gligoric \etal~\cite{Ekstazi:2015,Gligoric:2015} proposed Ekstazi, a firewall-based approach that selects test cases based on the changes to their dependent files.








\section{Conclusion and future work}
\label{sec:Conclusion}

In this paper, we present \toolName (Commit Adaptive Tool for Test suite Optimization), a tool implementing a TCS strategy for Java programs that selects TCs to be re-executed by comparing the call graphs of the two versions of the SUT (\ie the versions before and after a change).

\toolName encourages the Agile practice of continuous testing intercepting git commits and providing tests results directly within the IDE. This relieves the developer of the burden of constant context switching between development and testing. 

In a Lean context, this tool represents the application of Jidoka, a concept used in Lean Manufacturing that suggests to ``stop the line'' (intended is the production line) in case a problem is detected \cite{Womack2003}. Jidoka has the goal of automating quality assurance processes and to avoid unnecessary rework: it does this by stopping the entire production until it is clear how to fix an occurring problem. This strategy is not perfect for everyone, but in context where the cost of fixing an issue is higher than producing code, it is worth consideration. \toolName implements the concept of ``stopping the line'' in two ways: it automatically verifies (tests) the source code change (this cannot be deactivated) and---in case of a problem---warns the user. In the future, the strategies for regression testing will be extended---adding also test case prioritization, to increase the efficiency of the testing tool and to reduce testing time---as well as the way users are notified about the problem: from unobtrusive warnings to more aggressive strategy like to forbid a commit in case of a failing test case.

\balance 
\bibliographystyle{IEEEtran}
\bibliography{sample-bibliography.bib}

\end{document}